\documentclass[12pt]{iopart}

\usepackage[T1]{fontenc}
\usepackage{bm}
\usepackage{graphicx}
\usepackage{epstopdf}
\usepackage{float}
\usepackage[caption=false]{subfig}
\usepackage{ulem}
\usepackage[euler]{textgreek}
\usepackage{color}

\usepackage{mathrsfs}

\begin{document}

\title{Singlet and triplet trions in WS$_2$ monolayer encapsulated in hexagonal boron nitride}

\author{D Vaclavkova$^{1,2}$, J Wyzula$^{1,2}$, K Nogajewski$^{1,3}$, M Bartos$^{1}$, A~O~Slobodeniuk$^{1}$, C Faugeras$^{1}$, M Potemski$^{1,3}$, M R Molas$^{1,3}$}

\address{$^{1}$ Laboratoire National des Champs Magn\'etiques Intenses, CNRS-UGA-UPS-INSA-EMFL, 25, avenue des Martyrs, 38042 Grenoble, France}
\address{$^{2}$ Department of Experimental Physics, Faculty of Science, Palack\'y University, 17.~listopadu 12, 771 46 Olomouc, Czech Republic}
\address{$^{3}$ Institute of Experimental Physics, Faculty of Physics, University of Warsaw, ul.~Pasteura 5, 02-093 Warszawa, Poland}

\ead{maciej.molas@fuw.edu.pl}
\vspace{10pt}
\begin{indented}
	\item[]May 2018
\end{indented}

\begin{abstract}
	Embedding a WS$_2$ monolayer in flakes of hexagonal boron nitride allowed us to resolve and study the photoluminescence response due to both singlet and triplet states of negatively charged excitons (trions) in this atomically thin semiconductor. The energy separation between the singlet and triplet states has been found to be relatively small reflecting rather weak effects of the electron-electron exchange interaction for the trion triplet in a WS$_2$ monolayer, which involves two electrons with the same spin but from different valleys. Polarization-resolved experiments demonstrate that the helicity of the excitation light is better preserved in the emission spectrum of the triplet trion than in that of the singlet trion. Finally, the singlet (intravalley) trions are found to be observable even at ambient conditions whereas the emission due to the triplet (intervalley) trions is only efficient at low temperatures.
\end{abstract}

%
%
%
%
%

\section{Introduction\label{sec:intro}}

Charged excitons or trions, the three-particle complexes composed of an electron-hole ($eh$) pair 
(exciton) and an excess carrier (electron or hole), are relevant excited states which appear 
in interband optical spectra of $n$- or $p$-type semiconductor structures such as quantum wells\cite{cox1993,buhmann1995}
and/or atomically thin layers of semiconducting transition metal dichalcogenides (S-TMDs)\cite{jones}.
In reference to a negative trion, the case being considered here,  the two electrons involved in this
complex may be characterized by either antiparallel or parallel alignment of their spins, thus 
forming a singlet- or triplet-state trion. Most of works on trions refer to their singlet 
configurations in which they appear as bound states (generally, below the energies of neutral excitons). 
The observation of triplet trions is more elusive. If the two electrons of a trion originate from 
the same/single conduction band valley, the exchange interaction between electrons (Pauli repulsion) 
makes triplet trions to be the unbound states\cite{shield,hadas,astakhov}. The Pauli repulsion can 
be, however, thought to be significantly weakened in semiconductors with two valleys, such as S-TMD 
monolayers\cite{ramas,liu,kormanyos}, which have recently attracted considerable attention\cite{kai2013,li2018,koperski}.
Then, the triplet trions may involve two electrons with the same spin but each from a different
valley (each electron with a different quantum number, the valley pseudospin). When taking into account
the characteristic patterns of spin-orbit split bands, the observation of optically active trions 
in both singlet (two electrons from the same valley) and triplet (electrons 
from different valleys) configurations is favoured in the so-called "darkish" monolayers (see Refs~\cite{liu,kormanyos,koperski} 
for details),  WSe$_2$\cite{singh,courtade2017} and WS$_2$\cite{boulesbaa,plechingerTRION,molasNanoscale,jadczaknano},
with fundamental optically active transitions associated with the upper spin-split conduction band 
subbands.

In this work, we profit from an improved quality of the WS$_2$ monolayer encapsulated in hexagonal boron nitride (hBN) flakes,
in which we are able to resolve the photoluminescence (PL) signal from the singlet (intravalley) as 
well as triplet (intervalley) states of negative trions, and to investigate their properties. The energy 
separation between the singlet and triplet states is found to be close to 7~meV. The studied optical 
orientation appears to be more effective for the triplet than for the singlet trions and, surprisingly,
it is also more effective when the excitation power is increased. Upon an increase in temperature, a quick
disappearance of the triplet trion is observed while the singlet trion remains visible in the PL spectra
up to room temperature. 

\section{Methods\label{sec:methods}}

The investigated samples, shown schematically in the insets of Fig.~\ref{fig:fig_0}, 
contain WS$_2$ monolayers (MLs) placed on a Si/(320 nm)SiO$_2$ substrate or
encapsulated in hBN flakes and supported by a bare Si substrate. They were obtained by two-stage polydimethylsiloxane (PDMS)-based\cite{gomez} mechanical exfoliation of WS$_2$ and hBN bulk crystals 
purchased from HQ Graphene. The WS$_2$ MLs deposited 
on a Si/SiO$_2$ substrate and a bottom layer of hBN in hBN/WS$_2$/hBN heterostructures were created in the course of non-deterministic exfoliation.
The assembly of hBN/WS$_2$/hBN heterostructures was realized via succesive dry transfers of WS$_2$ MLs and capping
hBN flakes from PDMS stamps onto the bottom hBN layers of an appropriate thickness so as to benefit from
intereference effects increasing the observed intensity of emission lines in the PL spectrum. 

The \textmu-PL measurements were carried out under laser excitations provided by a laser diode
($\lambda_{exc.}$=515~nm, $E_{exc.}$=2.408~eV; non-resonant, above-bandgap excitation)
or a tunable dye laser adjusted at $\lambda_{exc.}$=565~nm ($E_{exc.}$=2.195~eV; near-resonant, slightly above the A-exciton
excitation\cite{molasNanoscale}). The samples were mounted on a cold finger of a continuous flow cryostat 
fixed on x-y motorized positioners. The excitation light was focused by means of a 50x 
long-working-distance objective with a 0.5 numerical aperture providing a spot of about 1~\textmu m
diameter. The PL signal was collected via the same microscope objective, sent through a 0.5~m 
monochromator, and then detected by a liquid nitrogen cooled charge-coupled device (CCD camera).
The polarization control of both the excitation and detected light was implemented with a set 
of polarization optics (including a linear polarizer, $\lambda$/2 and $\lambda$/4 wave plates) 
positioned in the laser and/or in the signal beam.

\section{Experimental results and discussion\label{sec:pl}}

\begin{center}
	\begin{figure}[t]
		\centering
		\includegraphics[width=0.8\linewidth]{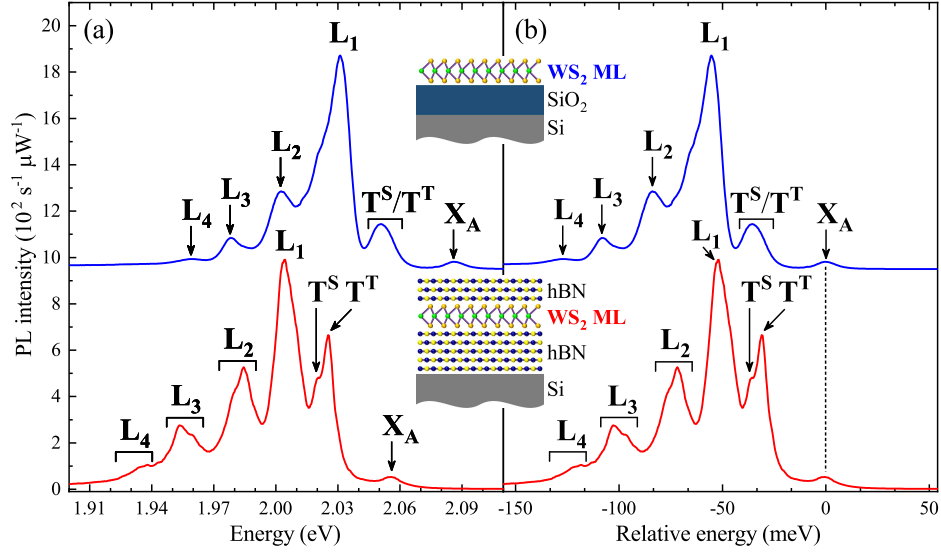}%
		\caption{Comparison between low-temperature ($T$=5~K) PL spectra measured on a WS$_2$ monolayer (blue curves) deposited on a Si/SiO$_2$ substrate and (red curves) encapsulated in hBN flakes, excited non-resonantly ($E_{exc.}$=2.408~eV and $P_{exc.}$=50~\textmu W) and plotted versus (a) absolute energy and (b) relative energy, $E-E_{\textrm{X}_\textrm{A}}$, where $E_{\textrm{X}_\textrm{A}}$ corresponds to the energy position of the neutral exciton, X$_\textrm{A}$, peak. The insets present the side-view schemes of the studied samples.}
		\label{fig:fig_0}
	\end{figure}
\end{center}

\begin{center}
	\begin{figure}[t]
		\centering
		\includegraphics[width=0.5\linewidth]{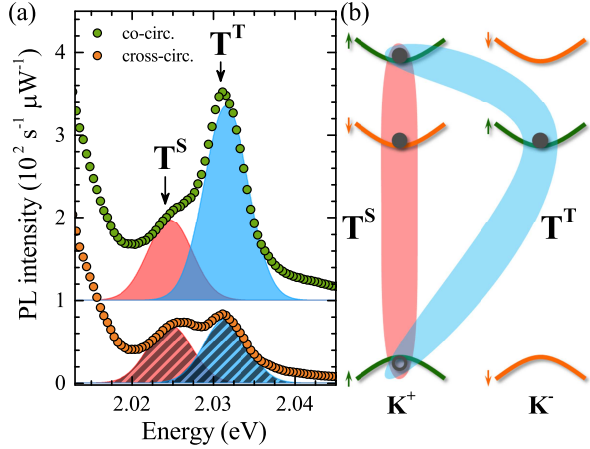}%
		\caption{(a) Helicity-resolved PL spectra of a WS$_2$ monolayer encapsulated in hBN flakes (at $T$=5~K) under near-resonant, circularly polarized excitation ($E_{exc.}$=2.195~eV and $P_{exc.}$=50~\textmu W). The red and blue Gaussians display fits to the corresponding helicity-resolved T$^\textrm{T}$ and T$^\textrm{S}$ lines. (b) Schematic illustration of possible spin configurations for negatively charged excitonic complexes: T$^\textrm{T}$ and T$^\textrm{S}$ trions formed at the K$^+$ point of the Brillouin zone, respectively.}
		\label{fig:fig_1}
	\end{figure}
\end{center}

A comparison between low-temperature PL spectra measured on the WS$_2$ monolayer
deposited on SiO$_2$/Si substrate (a "simple" structure) and on that 
encapsulated in hBN flakes (van der Waals heterostructure) is presented in Fig.~\ref{fig:fig_0}. 
Both spectra display several emission lines with a similar characteristic pattern 
already reported in a number of previous works on WS$_2$ monolayers in the "simple"~\cite{plechinger,shang,plechinger_nano,molas,koperski,molasSR,molasNanoscale,jadczaknano}
and van der Waals structures~\cite{wang2017}. In accordance with these reports, 
the highest energy emission peak (X$_\textrm{A}$) is attributed to the neutral 
exciton resonance. The following, lower-in-energy feature is assigned to a negative 
trion. A doublet structure (T$^\textrm{S}$ and T$^\textrm{T}$) of this feature 
is not resolved in our "simple" sample but becomes apparent in the van der Waals heterostructure. 
Following the previous observations\cite{boulesbaa,plechingerTRION,molasNanoscale}, 
the T$^\textrm{S}$ and T$^\textrm{T}$ emission peaks are identified as due to, 
correspondingly, singlet and triplet trions with two different configurations 
of electrons' spins. The ability to create charged excitons in the structures
under study most probably results from unintentional n-type doping of the WS$_2$ monolayer,
which is commonly reported for this material~\cite{plechinger,molasNanoscale,jadczaknano}.
Notably, the (optically active) exciton (X$_\textrm{A}$) 
in WS$_2$ monolayer is associated with the top spin-split valence band (VB) subband
and the upper spin-split conduction band (CB) subband. Thus, in our case of the 
monolayer with a low density of remote electrons, the singlet trion involves 
two electrons from the same valley whereas the triplet trion comprises two 
electrons from different valleys - see Fig.~\ref{fig:fig_1}(b) for a pictorial 
view of the intravalley singlet and intervalley triplet in a WS$_2$ monolayer.
A series of emission peaks labelled L$_1$, L$_2$, \ldots, in Fig. \ref{fig:fig_0} is 
repeatedly observed in PL spectra of WS$_2$ (and WSe$_2$) monolayers and attributed to 
recombination processes of "localized excitons"\cite{jones,plechinger,wangG2014,shang,arora,plechinger_nano,klopotowski,smolenski,molas,koperski,molasSR,jadczaknano,smolenski_WS2},
though a firm character of the L$_n$ emission series remain to be clarified in the future. 
Focusing on the neutral exciton peak X$_\textrm{A}$, we note its energy position at 
$E_{\textrm{X}_\textrm{A}}$$\sim$2.086~eV in our "simple" structure and at $\sim$2.056~eV in the 
heterostructure. Those values are related to the apparent amplitudes of the exciton 
binding energy $E_b$ and the bandgap energy $E_g$, ($E_{\textrm{X}_\textrm{A}}=E_g-E_b$). 
Bearing in mind different dielectric environments of WS$_2$ monolayers in our two samples,
we expect both $E_b$ and $E_g$ to be reduced in the heterostructure as 
compared to the case of "simple" structure (higher polarizability of the surrounding 
medium for a WS$_2$ monolayer encapsulated in hBN). A 30~meV red shift of the $E_{\textrm{X}_\textrm{A}}$
resonance in the heterostructure with respect to the "simple" structure indicates that 
the renormalization of $E_g$ gains over the renormalization of $E_b$. Most likely, 
however, this gain (30~meV) represents only a small fraction of the expected changes of $E_b$
(and $E_g$). In similar structures but with a WSe$_2$ monolayer,  $E_b$=290~meV was
estimated for a monolayer placed on Si/SiO$_2$\cite{Raja2017} and $E_b$=160~~meV 
was extracted for the monolayer encapsulated in hBN\cite{stierhBN}. Assuming a large (almost two times) 
change of  $E_b$ also in our WS$_2$ samples, a small difference ($\sim$3~meV) in the 
energetic separation between X$_\textrm{A}$ and T$^\textrm{S}$/T$^\textrm{T}$ peaks observed 
for these two samples is surprising. This separation energy, the trion binding energy, is often 
assumed to be proportional to the exciton binding energy\cite{stebe,thilagam}, thought this 
scaling rule may not be applied to the specific case of Coulomb interaction in an anisotropic dielectric medium~\cite{courtade2017,lin2014,komsa}.

In the following, we entirely focus on the  properties of negatively 
charged excitons (T$^\textrm{T}$ and T$^\textrm{S}$) 
in the encapsulated WS$_2$ monolayer. Their helicity-resolved 
PL spectra excited near-resonantly at an energy $E_{exc.}$=2.195 eV are presented in 
Fig.~\ref{fig:fig_1}(a). As schematically shown in Fig.~\ref{fig:fig_1}(b),
the appearance of trions with different spin configurations, $i.e.$ intravalley 
spin-singlet T$^\textrm{S}$ and intervalley spin-triplet T$^\textrm{T}$, 
is favourable in our sample, because the ground exciton state in monolayer WS$_2$ is 
optically inactive (dark)~\cite{molas,wang2017}.
Consequently, the neutral and charged exciton complexes are excited states of the system, 
because the electron forming an $eh$ pair is associated with the upper level in 
the CB~\cite{molasNanoscale}. An analogous situation takes place in the second 
tungsten-based ML, $i.e.$ WSe$_2$, which displays similar ordering of spin-orbit 
split subbands in the conduction band~\cite{molas,Zhou2017,Zhang2017,wang2017}.
In contrast, the bright MLs, $e.g.$ MoSe$_2$ and MoTe$_2$, are characterised by an
optically active (bright) ground-state exciton. This results in  
the observation of only the spin-singlet state of a charged exciton due to the Pauli exclusion principle.
To analyse trions' features in details, we fitted them with two Gaussian functions reflecting the T$^\textrm{T}$ and 
T$^\textrm{S}$ contributions, which fairly matches the 
experimental data (see Fig.~\ref{fig:fig_1}(a)).  
We found that the linewidths of the T$^\textrm{T}$ and T$^\textrm{S}$
peaks are of about 5~meV, which corresponds to the recently revealed narrowing effect 
of encapsulation in hBN associated with the suppression of mechanisms leading to inhomogeneous
broadening of excitonic emission lines~\cite{wang2017,cadizMoS2,wierzbowski2017}.

The energy separation, $\Delta_\textrm{S-T}$, between the T$^\textrm{T}$ and T$^\textrm{S}$
lines is found to be  $\Delta_\textrm{S-T}$=7$\pm$0.5~meV. This value is in very good agreement
with a recent assessment of  $\Delta_\textrm{S-T}$=8~meV from Raman coherence measurements~\cite{jakubczyk},
but somewhat smaller than the value ($\sim$11 meV) estimated from the deconvolution procedure of broad PL peaks~\cite{plechingerTRION}.
As recently discussed, the amplitude of  $\Delta_\textrm{S-T}$ is governed by Coulomb exchange between
the intravalley trion (both electrons in the same valley) and the intervalley trion
(electrons in two different valleys)\cite{YuH2014}. An additional electron
in the T$^\textrm{S}$ and T$^\textrm{T}$ complexes formed at the same K point of the Brillouin zone
occupies the same (opposite) valley for the T$^\textrm{S}$ (T$^\textrm{T}$) trion (a schematic representation of possible configuration
of trions is presented in Fig.~\ref{fig:fig_1}(b)).
Therefore, the coordinate parts of the wave-functions of these trions are different, $i.e.$ they belong
to different representations of the crystal symmetry group $D_{3h}$. It results in
unequal exchange energy corrections for the intra- and intervalley trions, respectively
(see Ref.~\cite{courtade2017} for details). Consequently, the degeneracy of the two spin 
configurations of the negative trion complex is lifted, $i.e.$ higher and lower energy states
are ascribed correspondingly to the intravalley (spin-singlet) and intervalley (spin-triplet)
charged excitons (see Fig.~\ref{fig:fig_1}(a)).

Next we would like to analyse in more detail the optical orientation of trions (helicity of the outgoing light with respect to the helicity of the excitation light). 
Both, T$^\textrm{T}$ and T$^\textrm{S}$ emission lines are more intense in the co-circular 
configuration of the excitation/emitted light as compared to the cross-circular one. This 
indicates a pronounced optically induced valley polarization of both excitonic 
transitions. Based on fitting the trions' contribution to the PL spectrum with a combination of two Gaussian curves, we were able to determine the circular
polarization degree of the emitted light defined as $\mathscr{P}=(I_{co}-I_{cross})/(I_{co}+I_{cross})$
for each trion component, where $I_{co}$ ($I_{cross}$) is the PL intensity detected in the circular 
polarization of the same (opposite) helicity as the excitation light. It is worth to point 
out that the mechanisms, which are responsible for preservation of excitonic complexes' polarization 
in S-TMD monolayers have been investigated and discussed in detail in several works, $e.g.$ \cite{mak2012,hanbicki2016,plechingerTRION}.
The T$^\textrm{T}$ line related to the spin-triplet trion displays the $\mathscr{P}$ value 
of almost 50$\%$, while the T$^\textrm{S}$ line ascribed to the spin-singlet
trion is characterized by a significantly smaller $\mathscr{P}$, of about 16$\%$. 
The extracted significant difference in the efficiency of optical orientation of both 
trion lines is similar to that already reported in Ref.~\cite{plechingerTRION}
(34$\%$ and 19$\%$, respectively) and indicates their different valley 
dephasing rates. According to Ref.~\cite{singh}, the fastest and 
hence the most efficient scattering mechanism responsible for quenching of the
polarization degree of charged excitons is due to scattering of an $eh$ pair
between valleys, which implies that the T$^\textrm{T}$ population feeds the T$^\textrm{S}$
one. The process is as follow: an $eh$ pair forming the T$^\textrm{T}$  trion
at the K$^-$ valley, shown in Fig.~\ref{fig:fig_1}(b), can scatter to
the opposite valley and form the T$^\textrm{S}$ trion at the K$^+$ point.
This process is mediated by the $eh$ exchange interaction towards an energetically 
favourable trion state. Other relaxation pathways, such as scattering of 
a given complex (T$^\textrm{S}$ or T$^\textrm{T}$) between valleys 
(a simultaneous transfer of three carriers, spin flips of two electrons 
and a hole from one valley to another, is required) or scattering between
T$^\textrm{S}$ and T$^\textrm{T}$ charged excitons formed at a given 
K$^\pm$ point (a spin flip of an excess electron is mandatory), are 
likely less efficient~\cite{singh}. Therefore, each of charged exciton complexes 
should be characterized by a different circular 
polarization degree, as it can be appreciated in Fig.~\ref{fig:fig_1}(a).
Worth noting is that the value of $\mathscr{P}$ for the neutral exciton 
($\sim$33$\%$, data not shown) is in our case similar to that previously reported (20$\%$ - 30$\%$)
in the experiments performed on "simple"  WS$_2$ monolayers~\cite{plechingerTRION,koperski,smolenski_WS2}.

\begin{center}
	\begin{figure}[t]
		\centering
		\includegraphics[width=0.5\linewidth]{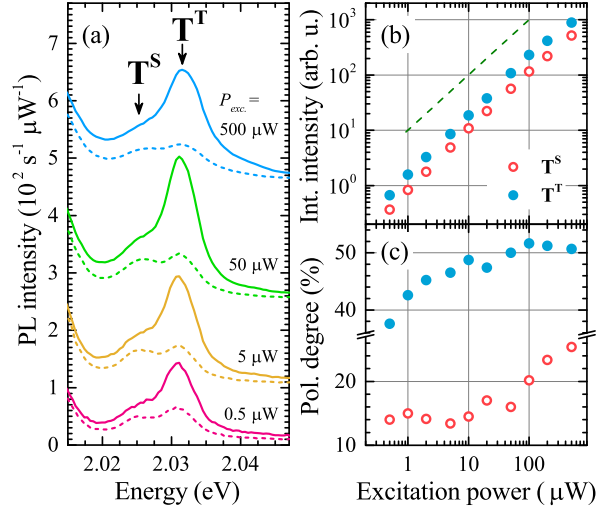}%
		\caption{(a) Power dependence of PL spectra of a WS$_2$ monolayer encapsulated in hBN flakes (at $T$=5~K) under near-resonant, circularly polarized excitation ($E_{exc.}$=2.195~eV and $P_{exc.}$=50~\textmu W). The intensities of the PL spectra are normalized by the excitation power. The solid and dashed curves display two configurations of circular polarization: co- and cross-circular, respectively. (b) Integrated intensities and (c) polarization degree of the intravalley and intervalley trions as a function of excitation power. The dashed green line in panel (b) is a guide to the eye indicating a linear increase.}
		\label{fig:fig_2}
	\end{figure}
\end{center}

To further discuss the properties of pronounced T$^\textrm{S}$ and T$^\textrm{T}$ 
features observed in the low-temperature PL spectra of our WS$_2$ monolayer
encapsulated in hBN flakes, we will focus on the evolution of the helicity-resolved spectra with the 
excitation power (see Fig.~\ref{fig:fig_2}). As illustrated in Fig.~\ref{fig:fig_2}(a),
both trions' lines can be recognised in a whole range of the
excitation power covering the change by 3 orders of magnitude (from 0.5~\textmu W to 500~\textmu W).
The integrated intensities of the T$^\textrm{S}$ and T$^\textrm{T}$ peaks
display a linear dependence on the excitation power (see Fig.~\ref{fig:fig_2}(b)),
as is expected for excitonic features\cite{Klingshirn}. Surprisingly, 
the polarization degrees of both the T$^\textrm{S}$ and T$^\textrm{T}$
trions, presented in Fig.~\ref{fig:fig_2}(c), grow 
with increasing excitation power. A double growth of the polarization degree
is seen for the intravalley T$^\textrm{S}$ trion, while for the intervalley 
T$^\textrm{T}$ complex it increases by $\sim$30$\%$ only. 
However, in terms of absolute $\mathscr{P}$ values, the increases are 
similar and amount to about 15$\%$. 
As clearly visible in Fig.~\ref{fig:fig_2}(c), 
there exists a distinct correlation between the changes in the polarization
degree of the T$^\textrm{S}$ and T$^\textrm{T}$ trions. Up to about 10~\textmu W,
the polarization degree of the T$^\textrm{T}$ complex increases
with the excitation power while that of the T$^\textrm{S}$ feature
essentially remains on the same level. Beyond 10~\textmu W, the situation 
inverts and now it is the T$^\textrm{S}$ trion whose polarization degree 
increases whereas that of the T$^\textrm{T}$ one progressively saturates 
at about 50$\%$. Worth noting is also a striking similarity between the 
shapes of the polarization degree versus excitation power dependences
recorded for the T$^\textrm{S}$ and T$^\textrm{T}$ trions - with respect 
to the 	center of the figure they present mirror images one to another. 
In order to understand these observations one 	has to realize that 
for a given helicity of the near-resonant excitation light (addressing 
either the K$^+$ or K$^-$ valley), the populations of both trion species 
in the 	complementary valley come from the intervalley scattering of 
photo-generated $eh$ pairs. In contrast to what happens in the valley 
addressed by the excitation light, where, as evidenced by measurements 
performed in the co-circular configuration, the T$^\textrm{T}$ complexes 
are being created roughly twice more efficiently than the T$^\textrm{S}$ ones, 
the probabilities of converting the scattered $eh$ pairs into T$^\textrm{T}$ 
or T$^\textrm{S}$ trions in the complementary valley are practically equal.
That is why there is no significant difference between the intensities of 
the T$^\textrm{S}$ and T$^\textrm{T}$ contributions to the PL spectrum recorded 
in the cross-circular configuration (see Fig.~\ref{fig:fig_1}(a) and 
dashed curves in Fig.~\ref{fig:fig_2}(a)). By combining these two facts
(approximately two times larger population of T$^\textrm{T}$ trions in the
valley addressed by the excitation light and roughly equal probabilities 
of creating T$^\textrm{T}$ and T$^\textrm{S}$ trions out of scattered eh
pairs) with the values of polarization degrees observed for the T$^\textrm{S}$
and T$^\textrm{T}$ complexes, one can estimate that at low excitation
power about 33$\%$ of photo-generated $eh$ pairs is subject to the
intervalley scattering. The convex (concave) shape of the polarization 
degree versus excitation power dependence for the
T$^\textrm{T}$ (T$^\textrm{S}$) trions (see Fig.~\ref{fig:fig_2}(c)) 
indicates that the probabilities of creating these trions out of scattered
$eh$ pairs are in fact not equal, but there exists some mechanism, 
which gives preference to the T$^\textrm{S}$ complexes. As can be seen, 
the excess amount of these trions decreases with increasing the excitation 
power. Last but not least, a general growth of the polarization degree 
with the excitation power observed for both trion species points at
direct or indirect excitation-power dependence of the intervalley
scattering efficiency of photo-generated $eh$ pairs. Indeed, by
making use of the same reasoning as presented above, one can
estimate that for the highest excitation powers shown in Fig.~\ref{fig:fig_2}(c) 
the amount of $eh$ pairs subject to the intervalley scattering is 
equal to about 28$\%$. This surprisingly small change in the
scattering efficiency has a profound effect on the polarization degree 
of both the T$^\textrm{S}$ and T$^\textrm{T}$ trions.
To the best of our knowledge an analogous effect
has not been reported so far in any S-TMD monolayer. When the excitation power
is increased, a number of formed $eh$ pairs grows and carriers 
start to occupy states characterized by larger $k$-vectors. Consequently, one can speculate
that scattering/relaxation processes become less efficient for states with larger $k$-vectors, which enables the polarization degree of trions to increase. This supposition, though conceivable, requires, however, a theoretical justification which stays beyond the scope of the present paper.

\begin{center}
	\begin{figure}[t]
		\centering
		\includegraphics[width=0.5\linewidth]{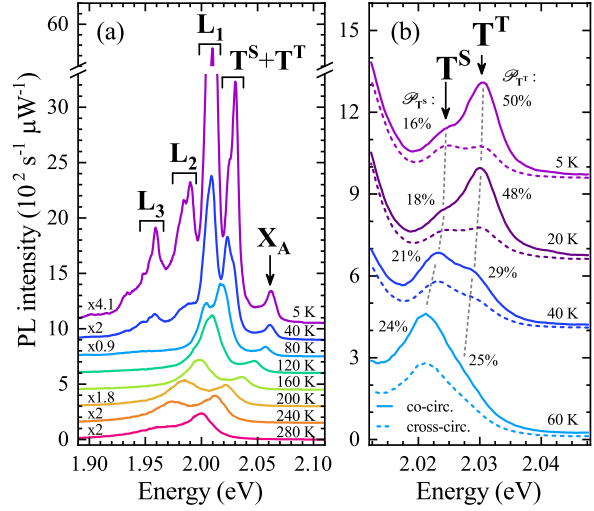}%
		\caption{(a) Temperature evolution of PL spectra measured on a WS$_2$ monolayer encapsulated in hBN flakes under near-resonant excitation ($E_{exc.}$=2.195~eV and $P_{exc.}$=20~\textmu W). (b) Temperature evolution of the corresponding PL spectra recorder under the same near-resonant excitation, but with the use of circularly polarized light. The spectra were detected in two circular polarizations of opposite helicities. The polarization degrees ($\mathscr{P}$) of individual trions are indicated above each pair of spectra. The energy range is narrowed to the emission of the T$^\textrm{S}$ and T$^\textrm{T}$ lines. The spectra are vertically shifted for clarity and some of them are multiplied by scaling factors in order to avoid their intersections with the neighbouring experimental curves or to make them better visible. The dashed lines in panel (b) are guides to the eye. }
		\label{fig:fig_3}
	\end{figure}
\end{center}

The temperature evolution of the PL spectra, excited with a moderate laser power,
is shown in Fig.~\ref{fig:fig_3}(a).  As often observed\cite{molasNanoscale,jadczaknano},
the low energy peaks associated with "localized excitons" (L$_1$, L$_2$, \ldots) 
are rapidly quenched with temperature and the neutral exciton emission dominates 
the spectra in the limit of high (room) temperature. The trion emission is also 
well visible, practically in a whole range of temperatures. Interestingly, each 
trion component displays very different behaviour. At low temperatures (up to 
around 20~K), the triplet (T$^\textrm{T}$) emission is much stronger than 
the singlet (T$^\textrm{S}$) one (see Fig.~\ref{fig:fig_3}(b)). On the other hand, 
the T$^\textrm{T}$ feature quickly disappears from the PL spectra with increasing 
temperature and only the T$^\textrm{S}$ peak is observed in the temperature range 
above 60~K. This behaviour might be accounted in terms of efficient thermal 
activation/ionization of the triplet-state trion, but somewhat less efficient 
activation of the singlet-state trion. This reasoning is in line with the fact 
that the binding energy of the T$^\textrm{T}$ trion state (energy distance to 
the X$_\textrm{A}$ exciton state) is smaller than the corresponding binding
energy of the T$^\textrm{S}$ trion state. Nevertheless, a simple scheme of 
thermal distribution of the T$^\textrm{S}$, T$^\textrm{T}$ and X$_\textrm{A}$ 
states fails to account for the observed temperature evolution of the emission 
spectra on a quantitative level. There must be some additional effects which 
imply blocking of the relaxation paths between these states, in particular, 
likely related to an inhibition of electron spin-flip transitions involved in the
T$^\textrm{T}$ and T$^\textrm{S}$ states. Regardless of complex physics which 
governs the non-thermal distribution of different excitonic complexes in our sample, 
which we are unable to clarify with the data in hands, it becomes apparent 
that the polarization degree of the T$^\textrm{S}$ state is growing with 
temperature in the range when the T$^\textrm{T}$ state progressively disappears 
from the spectra (see Fig.~\ref{fig:fig_3}(b)). Thus, in agreement with our 
previous arguments, the polarization degree of the T$^\textrm{S}$ trion is
largely governed by the appearance of the trion triplet states. We note, however,
that there must exist also other mechanisms which determine the efficiency 
of optical pumping in our samples. They are necessary to explain that starting 
from temperatures of about 60~K the polarization degree of the measured spectra
progressively vanishes, reaching practically a zero level at room temperature.

\section*{Conclusions}\label{sec:conclusions}
	
Trion's fine structure has been investigated in WS$_2$ monolayers embedded between two flakes of hexagonal
BN by means of helicity-resolved photoluminescence experiment. 
We found that at low temperatures the encapsulation of the WS$_2$ monolayer in hBN leads to the observation
of two well-resolved emission peaks due to two spin configurations of the negatively charged exciton: 
intervalley spin-triplet and intravalley spin-singlet. The magnitude of the splitting between them
is on the order of 7 meV and both of them posses an evidently different optically induced 
circular polarization. An increase in temperature brings a quick disappearance of the 
intervalley-related line from the PL spectra, while the emission of the intravalley trion is apparent 
up to room temperature.

\section*{Acknowledgements}

The work has been supported by the European Research Council (MOMB project no. 320590), the EC Graphene Flagship project (no. 604391), the National Science Centre (grants no. DEC-2013/10/M/ST3/00791, UMO-2017/24/C/ST3/00119), and the Nanofab facility of the Institut N\'eel, CNRS UGA. The support from the Fundation for Polish Science through the project "The Atomically thin semiconductors for future optoelectronics" carried out within the TEAM programme and co-financed by the European Union under the European Regional Development Fund is also acknowledged.

\section*{References}

\bibliographystyle{iopart-num}
\bibliography{biblio_hBN}

\end{document}